\documentclass[a4paper,fleqn,usenatbib]{mnras}

\usepackage{newtxtext,newtxmath}
\usepackage[T1]{fontenc}
\usepackage{ae,aecompl}

\usepackage[dvipsnames]{xcolor}

\usepackage{graphicx}	
\usepackage{amsmath}	
\usepackage{BibDef}

\usepackage{ragged2e}

\usepackage{booktabs}

\usepackage{multirow}

\usepackage{mathrsfs}

\usepackage{comment}

\usepackage[normalem]{ulem}

\newcommand{\msun}{\ensuremath{\, \mathrm M_{\sun{}}}}

\newcommand\numberthis{\addtocounter{equation}{1}\tag{\theequation}}

\title[MBHB hardening in rotating systems]{Stellar hardening of massive black hole binaries: the impact of the host rotation}

\author[L. Varisco et al.]{
Ludovica Varisco,$^{1}$\thanks{E-mail: l.varisco4@campus.unimib.it}
Elisa Bortolas,$^{1,2}$
Massimo Dotti,$^{1,2,3}$
and
Alberto Sesana$^{1}$
\\
$^{1}$Dipartimento di Fisica ``G. Occhialini'', Università degli Studi di Milano-Bicocca, Piazza della Scienza 3, I-20126 Milano, Italy\\
$^2$INFN, Sezione di Milano-Bicocca, Piazza della Scienza 3, I-20126 Milano, Italy\\
$^3$INAF, Osservatorio Astronomico di Brera, Via E. Bianchi 46, I-23807, Merate, Italy\\
}

\date{Accepted XXX. Received YYY; in original form ZZZ}

\pubyear{2021}

\begin{document}
\label{firstpage}
\pagerange{\pageref{firstpage}--\pageref{lastpage}}
\maketitle

\begin{abstract}
Massive black hole binaries (MBHBs) with masses of $\sim 10^4$ to $\sim 10^{10} \msun$ are one of the main targets for currently operating and forthcoming space-borne gravitational wave observatories. In this paper, we explore the effect of the stellar host rotation on the bound binary hardening efficiency, driven by three-body stellar interactions. As seen in previous studies, we find that the centre of mass (CoM) of a prograde MBHB embedded in a rotating environment starts moving on a nearly circular orbit about the centre of the system shortly after the MBHB binding. In our runs, the oscillation radius is $\approx$ 0.25 ($\approx$ 0.1) times the binary influence radius for equal mass MBHBs (MBHBs with mass ratio 1:4). Conversely, retrograde binaries remain anchored about the centre of the host. The binary shrinking rate is twice as fast when the binary CoM exhibits a net orbital motion, owing to a more efficient loss cone repopulation even in our  spherical stellar systems. 
We develop a model that captures the CoM oscillations of prograde binaries; we argue that the  CoM angular momentum gain per time unit scales with the internal binary angular momentum, so that most of the displacement is induced by stellar interactions occurring around the time of MBHB binding, while  the subsequent angular momentum enhancement gets eventually quashed by the effect of dynamical friction. 
The effect of the background rotation on the MBHB evolution may be relevant for LISA sources, that are expected to form in significantly rotating stellar systems.
\end{abstract}




\section{Introduction}\label{sec:intro}

In the past two decades massive black holes (MBHs) have been recognized as an integral component of the galaxy formation and evolution process \citep[e.g.][]{2006MNRAS.365...11C,2008ApJS..175..356H}. Dark massive compact objects (i.e. MBHs) have been observed to be ubiquitous in galaxy centers \citep[see][and references therein]{2013ARA&A..51..511K} and their black hole nature have been recently corroborated by the Event Horizon Telescope observations of the nucleus of M87 \citep{2019ApJ...875L...1E}.

In the hierarchical clustering scenario, these MBHs grow along the cosmic history together with their galaxy hosts, increasing their mass primarily via accretion of cold gas promoted by secular instabilities within the galactic potential and/or by mergers with other galaxies \citep[e.g.][]{2000MNRAS.311..576K,2003ApJ...582..559V}. In this scenario, following the merger of two galaxies each hosting a MBH, a MBH binary (MBHB) is expected to form \citep{1980Natur.287..307B}.
The dynamical evolution of MBHBs has received a lot of attention in recent years, owing to the possibility of revealing their gravitational wave (GW) signals with current pulsar timing array (PTA) experiments \citep{2016MNRAS.458.3341D,2016MNRAS.455.1751R,2019MNRAS.490.4666P,2020ApJ...905L..34A}, sensitive to MBHBs of $\sim 10^9\msun$ at $z<1$ 
\citep{2008MNRAS.390..192S}, and with the planned Laser Interferometer Space Antenna  \citep[LISA,][]{2017arXiv170200786A}, which will detect coalescing MBHBs with masses in the range $10^3\msun-10^7\msun$ anywhere in the Universe \citep{2016PhRvD..93b4003K}.

The `vanilla' evolution of MBHBs has been laid out already in \citep{1980Natur.287..307B}. In the aftermath of a galaxy merger, dynamical friction (against stars, gas and dark matter) efficiently brings the two MBHs hosted by the parent galaxies to the center of the merger remnant. When the two MBHs feel each other potential, they form a bound binary which responds to the collective torque of the large scale distribution of matter as a single object, making dynamical friction inefficient. For typical MBHs of $10^6\msun-10^9\msun$, this occurs at $\sim 1-10$ pc, whereas GW emission can only drive the system to coalescence in less than an Hubble time from a separation of few milliparsecs \citep[e.g.][]{2007ApJ...660..546S}. The bridging of the three order of magnitude gap in between goes under the name of final parsec problem \citep{Milosavljevic03}, and its solution relies on the local interaction of the binary with its immediate dense surrounding of stars and gas \citep[see][for a review]{2012AdAst2012E...3D}.

Since the 90's it has been realized that three-body interactions between the MBHB and stars intersecting its orbit can efficiently extract energy and angular momentum from the binary: which is known as slingshot mechanism \citep{1992MNRAS.259..115M,1996NewA....1...35Q}. Shrinking the binary by orders of magnitudes to prompt a GW driven coalescence, however, requires to supply the system with a mass in stars which is several time larger than its own mass \citep{2005LRR.....8....8M} in a `cosmologically short' timescale (i.e. $\lesssim $ Gyr). Since stars interacting with the MBHB are expelled from the core of the galaxy, the coalescence of MBHBs require an efficient mechanism to repopulate stars on orbits intersecting the binary path, i.e. the binary loss cone. In spherically symmetric stellar systems the loss cone repopulation relies on two-body relaxation, and for typical galaxies it turns out to be of the order of the Hubble time or longer \citep{2001ApJ...563...34M}. 

This observation has triggered both (semi)analytical \citep[e.g.][]{2004ApJ...606..788M,2013ApJ...774...87V} and numerical \citep[e.g.][]{2006ApJ...642L..21B,2011ApJ...732L..26P,2011ApJ...732...89K,2013ApJ...773..100K,2017MNRAS.464.2301G, Bortolas2018b} investigations of MBHB evolution in more realistic stellar systems, including flattening, traxiality and rotation, which are expected as a result of the merger of the progenitor galaxies \citep[e.g.][]{Bortolas2018}. The general consensus emerging from this body of work is that the bulge resulting from a galaxy merger has enough level of triaxiality to allow loss cone repopulation in a relatively short timescale,\footnote{This is because in a triaxial potential individual orbits do not preserve their angular momentum and can diffuse into the loss cone over timescales which are much shorter than two body relaxiation time.} leading to final coalescence on timescales of Gyrs or less \citep{2015ApJ...810...49V}.

Besides geometry (sphericity, axisimmetry, triaxiality), another fundamental property of a stellar bulge that can influence the evolution of the hosted MBHB is net rotation. It is in fact known that retrograde stars extract more efficiently angular momentum leading to eccentricity growth, whereas prograde stars promote circularization\footnote{Assuming a cartesian reference centered in the MBHB center of mass, and the binary orbiting in the $x-y$ plane, a prograde (retrograde) star has the $z$ component of its angular momentum aligned (antialigned) to the MBHB angular momentum.} \citep{Sesana11}. Moreover, a MBHB embedded in retrograde stellar systems secularly change its orbital plane to align its orbital angular momentum to that of the stars \citep{Gualandris12}. The importance of these findings stem from the fact that GW emission is much more efficient in eccentric binaries \citep{1964PhRv..136.1224P}, thus significantly reducing MBHB merger timescales. Moreover, LISA will have the capability of measuring the MBHB eccentricity \citep{2016PhRvD..94f4020N}, thus providing important information in the reconstruction of the dynamical processes driving the pairing and hardening phase.

The aforementioned early results have been subsequently more rigorously formalized in \cite{2017ApJ...837..135R} and extensively investigated numerically in \cite{Holley-Bockelmann15, Mirza17} and \cite{Khan20}. These latter works found that the center of mass (CoM) of a prograde binary within rotating systems does not stay put in the centre (except for the traditional Brownian motion that was already studied e.g. by \citealt{Merritt01, Chatterjee03, Milosavljevic03, Bortolas16}), but starts to move in approximately circular orbits around the CoM of the stellar system. Contextually, the binary is found to shrink more effectively. Since in those simulations the stellar system is also flattened by rotation, it is not clear whether those effects are purely induced by rotation, and their physical origin has not been  investigated in depth. 

In this paper we perform a detailed study of the wandering of the MBHB CoM in a rotating stellar system. By means of controlled N-body experiments that keep the shape of the stellar distribution spherically symmetric while introducing net rotation, we isolate the role of rotation in the dynamical evolution of the MBHB CoM and build a sound analytical model that describes the outcome of the simulations. The paper is organized as follows. The setup of our N-body experiments is described in Section \ref{sec:methods} and the resulting MBHB CoM evolution is presented in Section \ref{sec:results} and modeled analytically in Section \ref{sec:model}. Finally, we discuss the relevance of this physical mechanism for real-life astrophysical systems in Section \ref{sec:discussion}.


\section{Simulations setup}\label{sec:methods}
In order study the effects of the system rotation on  the evolution of  MBHBs, we chose to initialize the host system as a spherically symmetric distribution of stars. This allows us to isolate the effect of the system rotation from  the impact of galaxy morphology, thus preventing the MBHB evolution to be affected by the combined effect of both rotation and deviation from spherical symmetry. The host system is first initialized following an \citet{Hernquist90} density profile:
\begin{equation}
\rho(r)= \frac{M_{\rm tot}}{2\pi} \frac{r_0}{r}\frac{1}{(r_0+r)^{3}} 
\end{equation}
with total mass of stars $M_{\rm tot}$, inner density slope $\gamma=1$ and scale radius $r_0$. We set our model units (MU) such that $M_{\rm tot}=G=r_0=1$, with $G$ gravitational constant. 

The stellar velocities are initialized at equilibrium in the potential well generated by the stellar distribution itself and by a primary massive black hole (MBH) of $M_{\bullet}=0.005 \, M_{\rm tot}$, at rest in the origin of the system.

We introduced rotation in our model following the same procedure adopted by \cite{Khan20}, i.e. by flipping the $z-$component of the angular momentum ($L_z$) of particles with initially negative $L_z$, for the co-rotating cases, and flipping those with positive initial $L_z$, for the counter-rotating case. 
In principle we could initialize a flattened system with a morphology directly linked to the degree of  rotation by sampling a distribution function of the form $f(E,L_z)$, as done, e.g., in \cite{Wang2014}. We however decided to enforce the spherical symmetry of the stellar spatial distribution, to isolate the effect of rotation only, as clarified above.
A secondary MBH is introduced in the system at an initial separation of $ r_0$ with initial tangential velocity equal to $70 \%$ the circular velocity at $r_0$ and with null radial velocity. 
In all simulations, the angular momentum  of the MBH pair is initially perfectly aligned (or anti-aligned, for the counter-rotating case) with the system angular momentum.

\begin{table}
	\centering
	\caption{Parameters of the simulations presented in this work. The model names have been chosen as follows: the capital letter `P' refers to prograde rotators while `R' refers to the retrograde rotators, the number indicates  number of particles of the simulation ($1$ for $N=256$ k particles, $2$ for $N=512$ k particles and $3$ for $N=1$ M particles); finally, the letter `e' refers to equal mass binaries ($q=1$) while `u' indicates unequal mass binaries ($q=0.25$). See the text for more details.}
	\label{tab:run_params}
	\begin{tabular}{cccc} 
		\hline
		Model & $N$ & $q$ & Rotation \\
		\hline
		P1e & $256$ k & $1$ & co-rotating \\
		P1u & $256$ k & $0.25$ & co-rotating \\
		P2e & $512$ k & $1$ & co-rotating \\
		P2u & $512$ k & $0.25$ & co-rotating \\
		P3e & 1 M & $1$ & co-rotating \\
		P3u & 1 M & $0.25$ & co-rotating \\
		R2e & $512$ k & $1$ & counter-rotating \\
		AP3e & 1 M & $1$ & co-rotating, anchored\\
		\hline
	\end{tabular}
\end{table}

We performed a suite of direct summation N-body simulations varying the mass resolution (i.e. the total number of particles $N$) and the binary mass ratio $q\leq1$ ($q=1, 0.25$). The simulations initializing parameters are summarized in Tab.~\ref{tab:run_params}. The labels of the runs are assigned so that the trailing capital letter refers to whether the (spherical) host system rotation is prograde (`P') or retrograde (`R') with respect to the MBHB initial orbit; the subsequent number indicates the number of particles in the simulation (1 for $N=256$~k, 2 for $N=512$~k and 3 for $N=1$~M); finally, the letter `e' refers to equal mass MBHs ($q$ = 1) while ‘u’ indicates unequal mass MBHs ($q = 0.25$). Note that the parameters of run P3e and P3u are similar to the runs $P_{1.00}$ and $P_{0.25}$ in \cite{Khan20}. In particular, the aforementioned runs present the same total number of particles ($N=1$M) and the same MBH mass ratios ($q=1$ and $q=0.25$, respectively). However, it is important to remember that the main difference of our work with respect to \cite{Khan20} consists in the different geometry of the host system: while \cite{Khan20} study the evolution of a MBHB in a rotating flattened Dehnen profile (with $\gamma=1$ and with a minor to major axis ratio of $0.8$), we study how a MBHB evolve in a spherical rotating stellar system. This is because we are interested in investigating the effect of the pure net system rotation on the MBHB evolution and hardening, and the introduction of a flattening would entangle the interpretation of our results.

We additionally performed a simulation with the same parameters as the P3e model (i.e. the highest resolution simulation with equal-mass binary co-rotating with the spherical stellar distribution) in which we forced the binary to stay anchored in the center of the system; we labelled this run as AP3e. More specifically, we took the snapshot at time  $t = 30.375$ (shortly after the formation of the bound binary): at this time we restarted the run forcing the binary centre of mass to sit at the centre of the system. Every $\Delta t=1.5625\times10^{-2}$ we  recursively computed   the centre of mass position and velocity of all particles (excluding the MBHs) within 2.35$r_0$, which roughly coincides with the half mass radius of the system.\footnote{The recentering is performed 5 times per step, with the binary centre of mass as the initial guess.} Then, we set the centre of mass position and velocity of the binary equal to the aforementioned one for the entire duration of the run. Note that the recentering significantly slowed down the integration: for this, AP3e was only evolved for $t\approx$  45 time units after the restart, while all other runs were evolved for at least 160 time units.

The initial conditions were evolved using the direct-summation N-body code \textsc{HiGPUs}, designed to run on GPU accelerators. \textsc{HiGPUs} features a very accurate, sixth order Hermite scheme with block time-steps \citep{Capuzzo-Dolcetta13}. 
The computation of the timestep is performed by combining the fourth and sixth order Aarseth criterion \citep{Aarseth03, Nitadori08}, with the respective accuracy parameters equal to 0.01, 0.45.
We set the softening parameter $\epsilon=10^{-4}$ for star-star interactions, $\epsilon=10^{-6}$ for MBH-MBH interactions, while the softening for mixed stellar-MBH interactions is set equal to the geometric average of the two.
For a typical run with $1$M particles, evolved for $\approx 200$ time units, the wall clock time needed is $\approx 110$ hours, using one node equipped with two  NVIDIA TeslaTM V100 GPUs, and four cpu cores.  

\section{Results}
\label{sec:results}

\begin{table}
	\centering
	\caption{For each run, the binary CoM radius is averaged over the time interval from $t=75$, where all models have settled around a nearly constant value, to $t=175$. The binary influence radius is computed using the definition in Eq. \ref{eq:Rinf} and averaged over the same time interval of $R_{\rm b}$, while the Brownian radius is computed via Eq.~\ref{eq:r_Brown}, as better detailed in the text.}
	\label{tab:Rbin}
	\begin{tabular}{cccc} 
		\hline
		Model & Binary CoM & Binary influence & Binary Brownian \\
		&  final radius (MU) &  radius (MU)  &  radius (MU) \\
		\hline
		P1e & 0.047 & 0.22 &  0.011\\
		P1u & 0.020 & 0.16 &  0.014\\
		P2e & 0.058 & 0.22 &  0.008\\
		P2u & 0.012 & 0.16 &  0.010\\
		P3e & 0.065 & 0.22 &  0.006\\
		P3u & 0.026 & 0.15 &  0.007\\
		R2e & 0.010 & 0.20 &  0.008\\
		\hline
	\end{tabular}
\end{table}

\subsection{Evolution of the orbital parameters}

\begin{figure}
   \centering
   \includegraphics[width=\hsize]{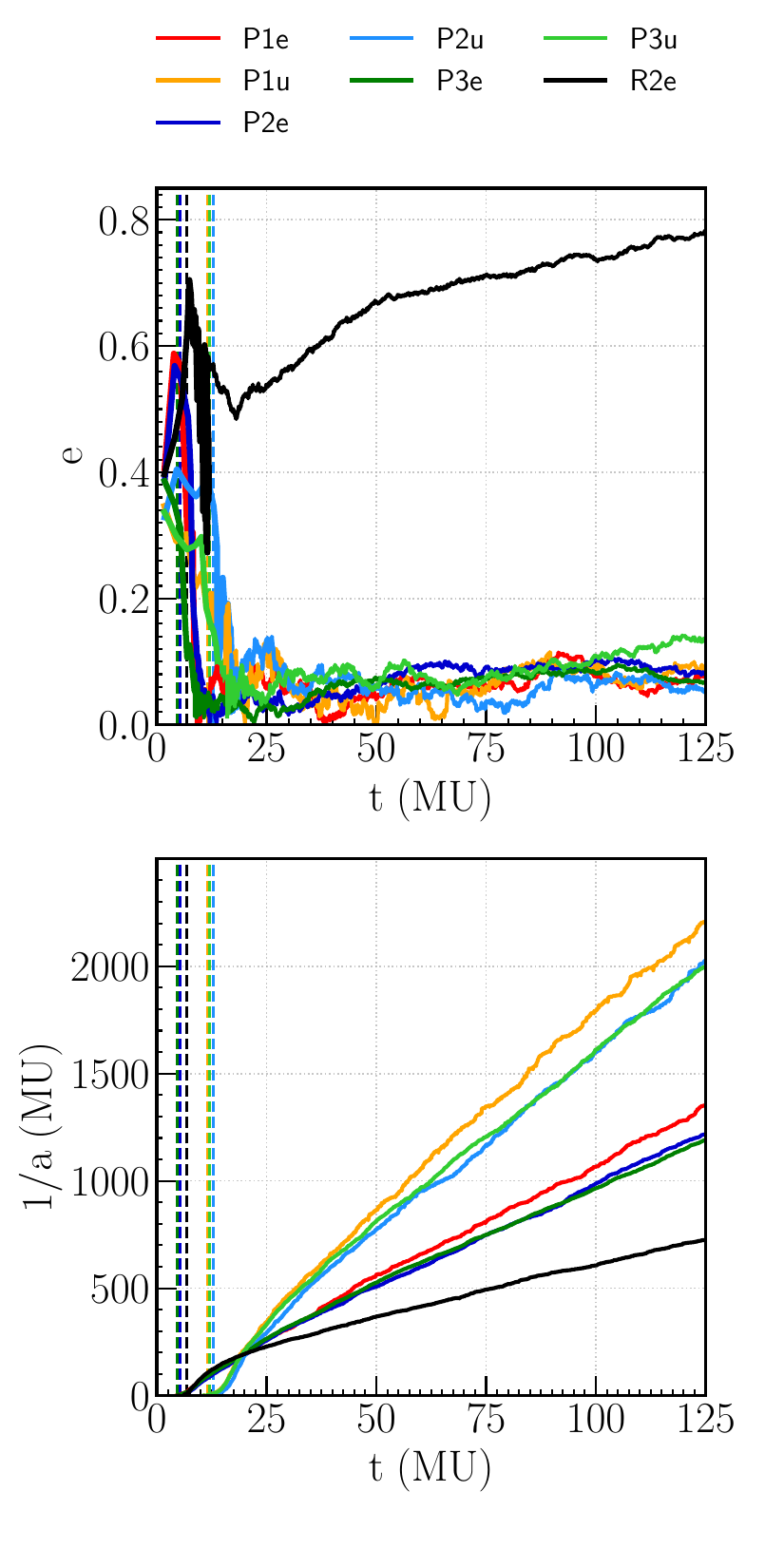}
      \caption{Time evolution of eccentricity (upper panel) and inverse semi-major axis (bottom panel) for each simulation. Note that prior to the binary formation time (indicated with a vertical dashed line) the binary orbital parameters are computed  via Eq.~\ref{eq:periapo}, while the standard Keplerian parameters are shown after the binary formation time.
              }
         \label{bh_prop}
\end{figure}

Fig.~\ref{bh_prop} shows the evolution of the MBHB properties as a function of time, and specifically the binary eccentricity $e$ and the  inverse of its semi-major axis, $1/a$. 
The dashed vertical lines indicate the binary formation time $t_{\rm bf}$, chosen as the instant at which a bound Keplerian binary forms. Note that the eccentricity and semimajor axis are computed as the standard Keplerian parameters from the binary formation time. Prior to that, these quantities are evaluated as:
\begin{equation}\label{eq:periapo}
\begin{split}
a_{\rm unbound} &=\frac{r_a+r_p}{2}\\
e_{\rm unbound} &= \frac{r_a-r_p}{r_a+r_p}
\end{split}
\end{equation}
where $r_p$ and $r_a$ respectively represent the pericentre and apocentre separations, computed once for each complete radial oscillation.

\begin{figure}
   \centering
   \includegraphics[width=\hsize]{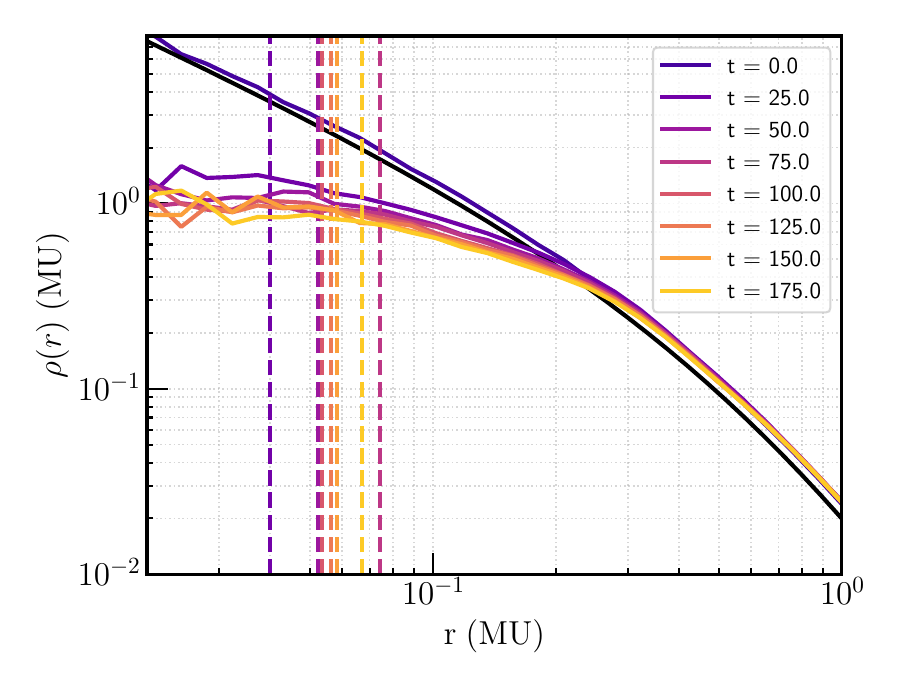}
      \caption{Evolution of the stellar density profile at different simulated times for model P3e. Density profiles are drawn at six different times: from the dark purple line at $t = 0$ to the yellow line at $t = 150$. Each profile was obtained averaging over five subsequent time-steps. The black solid line is the initialized theoretical \citeauthor{Hernquist90} profile. The vertical dashed lines, with the same color code of the density profiles, indicate the binary CoM radius at the corresponding time. The position of the binary CoM is not shown for $t=0$ since a bound binary has not formed yet. It is evident that, even though at larger radii all the profiles are consistent with the model, the central density is progressively reduced with time as an effect of slingshot interactions; the binary CoM always remains within the carved, almost constant density inner region.
              }
         \label{densityprofile}
\end{figure}

Fig.~\ref{bh_prop} shows the different stages characterizing the MBHs orbital evolution. 
Initially, the MBHs mutual separation is reduced via dynamical friction \citep{Chandrasekhar43}.
In our models, the two MBHs are initially placed at a relatively small  separation, thus this  phase is very short, and it ends roughly with the formation of a bound binary. 
When the binary reaches a separation comparable to the MBHB influence radius, defined as the radius of a sphere containing twice the MBHB mass in stars:
\begin{equation}\label{eq:Rinf}
M_*(r < r_{\rm inf}) = 2M_b
\end{equation}
three body scatterings with stars start to efficiently extract energy and angular momentum from the binary, adding up to the effect of dynamical friction and excavating a core in the stellar density profile \citep[e.g.][]{Milosavljevic03,Sesana08}; 
the scouring of the density profile in time is shown in Fig.~\ref{densityprofile} for model P3e.
The MBHB eventually reaches the hard binary separation $a_h$, i.e. the separation at which the binary binding energy exceeds the kinetic energy of the field stars:
\begin{equation}
a_h=\frac{GM_2}{4 \sigma_*^2}    
\end{equation}
where $M_2$ is the mass of the secondary MBH and $\sigma_*$ is the velocity dispersion of field stars. At this stage, the binary hardening occurs by  stellar interactions only, and the binary hardens at a slower pace, until it reaches the separation at which GWs start to dominate its evolution.\footnote{Note that the integrator implements a purely Newtonian approach and the GW phase cannot be followed in the current setup.}

Fig.~\ref{bh_prop} shows that the dynamical friction (DF) driven inspiral is more efficient for equal mass binaries, as the intruding MBH has a larger mass. After the binary formation, the binary tends to circularize in all the prograde models. In the retrograde rotators instead the binary eccentricity follows a significantly different trend: after a short phase of slow decrease, $e$ starts rising and it reaches $e \simeq 0.8$ by the end of the run. This result is aligned with what found in previous studies addressing the binary eccentricity evolution in rotationally supported systems \citep[e.g.][]{Gualandris12}  in which the perturber interacts with stars with a net tangential (prograde or retrograde) motion.
The evolution of the inverse semi-major axis, showed in the lower panel of Fig.~\ref{bh_prop}, is an important measure of the binary energy change as a function of time.
All the simulated models follow a similar qualitative evolution: once the binary forms, the inverse semi-major axis undergoes a short phase of fast increase after which it increases almost linearly with time. As expected, the models with lower mass-ratio show a  faster binary shrinking compared to the corresponding equal mass case \citep{2006ApJ...651..392S}. 

In all runs, the slight dependence of the shrinking efficiency on the total number of particles may be at least partially ascribed to two-body relaxation, which refills the binary loss cone more efficiently for the less resolved runs. We would like to stress once more that, in our runs, the idealized assumption of spherical symmetry in the mass distribution  is made in order to isolate the impact of the system rotation  on the binary shrinking rate; deviations from sphericity would tangle the interpretation of our results, as global gravitational torques induced by a non-spherical morphology would non-trivially  impact the evolution of the binary hardening;  the impact of rotation and axisymmetry combined have been investigated in \citet{Holley-Bockelmann15, Mirza17, Khan20}.
It is important to note  that the counter-rotating case shows a significantly lower binary hardening compared to all the co-rotating models. This aspect is better dissected in the sections below.

\subsection{Center of mass evolution}

In line with previous literature on the topic \citep{Holley-Bockelmann15, Mirza17, Khan20}, we found that the binary CoM in the prograde runs starts moving on a nearly circular orbit about the centre of the system shortly after the binary formation time.  In this section, we investigate such behaviour in detail.
In order to characterize the binary CoM motion we first need to define a reference centre of the host stellar system. 
To define the system centre we proceed as follows. As a first guess we set the system CoM to coincide with the binary CoM. We proceed computing the CoM of the stars contained within a radius of $2.35 \, r_0$ and then re-centering the whole system at that position. The iteration is repeated five times per snapshot. 

All our results are presented in a reference frame centered in the above defined position.\footnote{Note that the strategy described here to find the centre of the stellar distribution is  the same used for  anchoring the binary at runtime for run AP3e. In addition, we explored another possibility for computing the centre of  the system:   we recursively computed the CoM of particles in a shrinking sphere whose maximum (minimum) radius was set to 100$r_0$  (1.5$r_0$); the radius was halved at each iteration. We found a very good match between the two described centering strategies, with mismatches much smaller that the wandering radius $R_b$.}

\begin{figure}
  \centering
   \includegraphics[width=\linewidth]{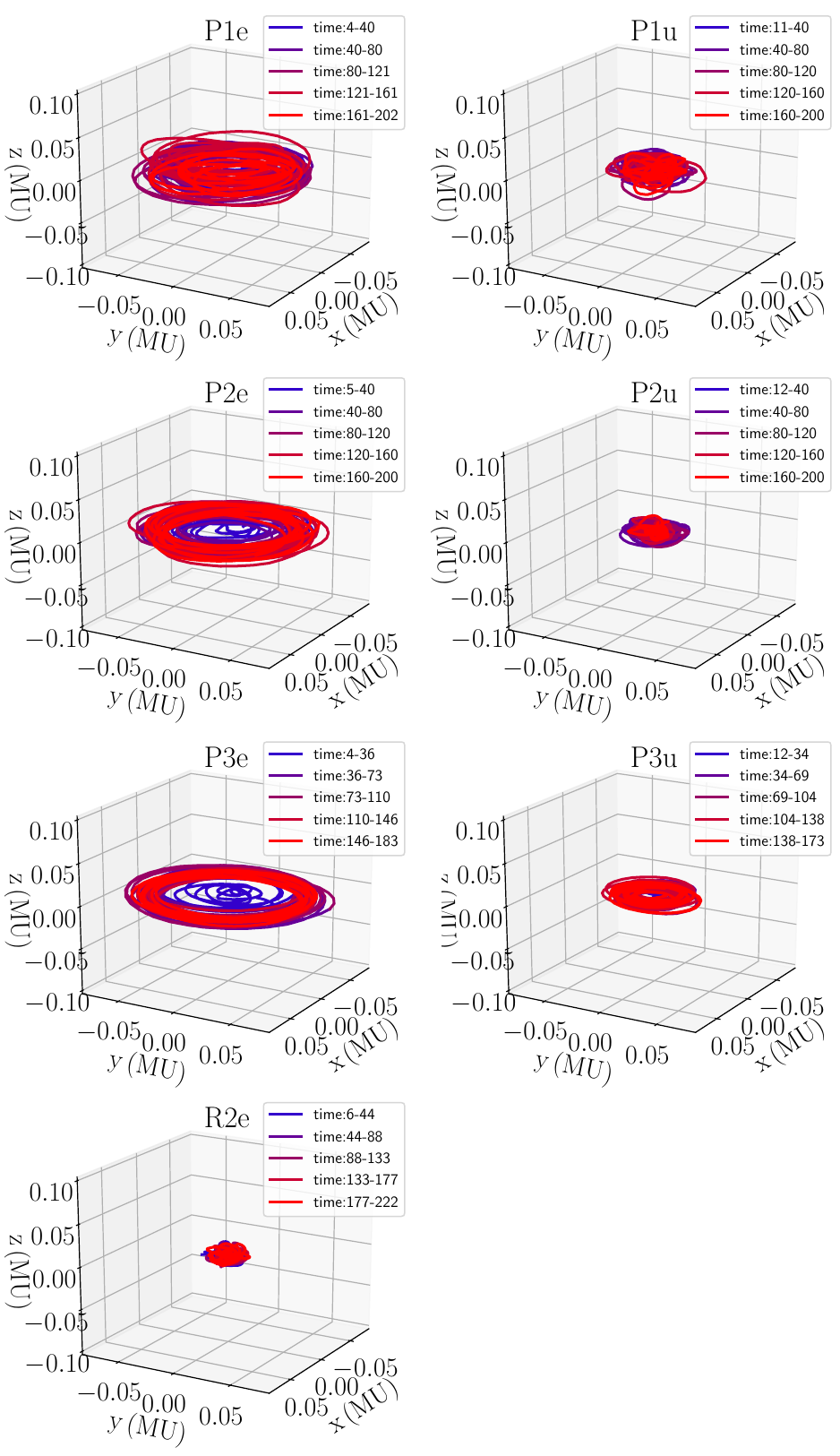}
      \caption{The figures show the three dimensional evolution of the  MBHB CoM trajectory for each of the runs presented in the current study. In each panel, the color code maps different time intervals in the orbital evolution, as shown in the legend. The initial time corresponds to the instant at which a bound Keplerian binary forms. 
              }
    \label{3d_trajectory}
\end{figure}

\begin{figure}
  \centering
  \includegraphics[width=\hsize]{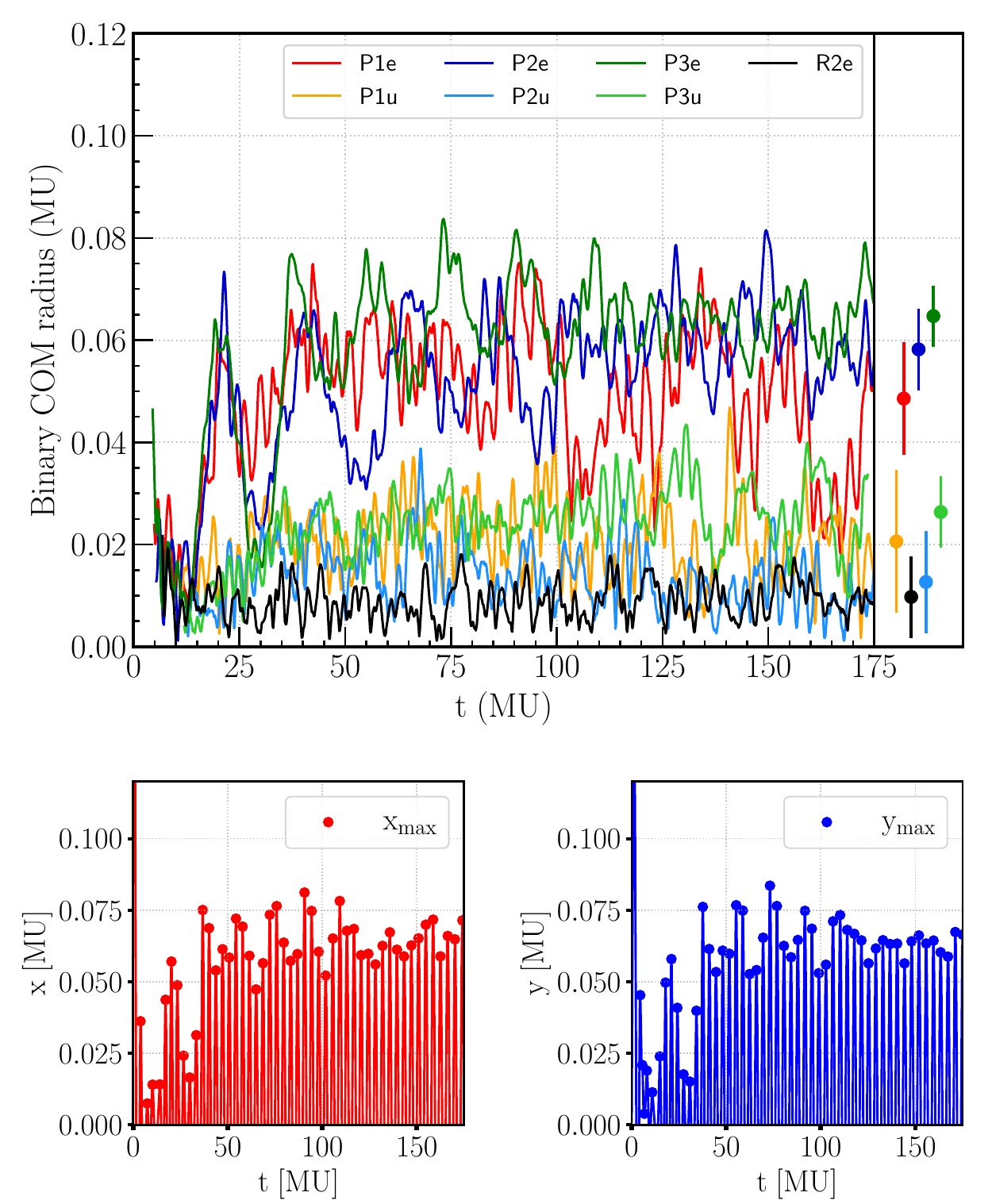}
    \caption{\textit{Upper panel}: on the left is shown time evolution of the MBHB CoM radius $R_b$ for the different runs presented in the paper; on the right, the dots indicate, for each run, the value of the binary CoM radius averaged between $t=75$ and $t=175$, while the error bars show the amplitude of the Brownian wandering radius (see Tab.~\ref{tab:Rbin}). \textit{Bottom panels}: on the left is shown the time evolution of the binary CoM orbit in the x-coordinate for run P3e, the dots indicating the local maxima. The analogus is shown on the right panel for the orbit in the y-coordinate.
              }
    \label{R_t}
\end{figure}

Fig.~\ref{3d_trajectory} shows the 3-dimensional trajectory of the MBHB CoM for all the simulated models. The top panel in Fig.~\ref{R_t} reports the temporal evolution of the distance between the MBHB CoM and the host centre ($R_{\rm b}$) after the binary formation time.  For co-rotating models, soon after the binary formation time $t_{\rm bf}$, the MBHB CoM starts orbiting the host centre with a rapidly increasing $R_{\rm b}$. After just few tens of time units the CoM settles on a nearly stable orbit. In particular, equal mass binaries show a faster rise of the CoM radius compared to the lower mass ratio cases. Moreover, the higher the binary mass ratio, the larger the final orbital radius: the two differ by nearly a factor 2. The retrograde run does not show the same behaviour, and the binary CoM remains very close to the centre, only experiencing the traditional Brownian wandering (as detailed below).  
Table \ref{tab:Rbin} reports the mean value of the final CoM radius for each model, computed averaging $R_{\rm b}$ over the time interval from $t=75$, where all models have settled around a nearly constant value, to $t=175$, along with the binary influence radius, $R_{\rm inf}$, averaged over the same time interval. 
Bottom panels of Fig \ref{R_t} show the time evolution of the binary CoM orbit in the x and y-coordinate (left and right panel, respectively) for the run 3Pe, thus pointing out the quasi-periodicity of the binary CoM orbit.
In co-rotating runs hosting equal-mass binaries the influence radius is $R_{\rm inf} =0.22$ while for co-rotating unequal-mass binaries is $R_{\rm inf} \simeq 0.16$. This difference is, at least partially, due to the different total mass of the MBHB ($M_{\rm b}=0.01$ if $q=1$, $M_{\rm b}=0.00625$ for $q=1/4$). The binary CoM oscillation in the prograde runs is much larger than the binary separation (see e.g. the values of $1/a$ in Fig.~\ref{bh_prop}), but smaller than the MBHB influence radius by a factor $3-5$ for the equal mass and by a factor $6-13$ for the unequal mass cases.

Note that the binary CoM oscillation found in the prograde runs is different than the traditional MBHB Brownian motion (see e.g. \citealt{Merritt01,Chatterjee03,Milosavljevic03, Bortolas16}). The latter is caused by the fact that slingshot ejections of stars with isotropic velocities w.r.t. the binary CoM induce a recoil in the binary CoM in random directions. The  associated displacement is contrasted by the effect of dynamical friction onto the binary as a whole: These two phenomena balance each other and result in a small and non-coherent wandering of the binary CoM, which however does not exhibit, on average, any net  angular momentum. The typical scale of the traditional Brownian wandering is smaller than the oscillation radius we find in prograde runs. In fact, the Brownian wandering radius scales as 
\begin{equation}\label{eq:r_Brown}
    r_{\rm Brown}\propto (m_\star/M_{\rm b})^{1/2}
\end{equation}
where $m_\star$ is the typical particle mass in the run and $M_{\rm b}$ is the binary total mass \citep{Merritt01}. \citet{Bortolas16} report a value of $r_{\rm Brown}\approx0.008$ for  $m_\star/M_{\rm b}\approx 2\times10^{-4}$ in a system whose initializing properties are analogue to the ones considered in the present work (i.e. an initial \citeauthor{Hernquist90} profile with unitary scale radius and total mass). By rescaling this value via Eq.~\ref{eq:r_Brown} we can infer the magnitude of the Brownian wandering in our runs: the computed values are shown in the left-hand column of Tab.~\ref{tab:Rbin}, and as error-bars in the  upper-right panel of Fig.~\ref{R_t}.
The Brownian radius is significantly smaller than the oscillation radius for prograde runs with the best adopted resolution, especially for the equal mass cases. The binary CoM displacement found in the retrograde case is instead compatible with being caused by the traditional Brownian motion. It is reasonable to interpret the trends shown in the upper panel of Fig.~\ref{R_t} for prograde runs  as the combination of the net rotation of the binary CoM, induced by the system rotation, and the traditional Brownian motion, that is likely responsible for at least part of the noise in the plotted curves. This idea is supported by the fact that the runs featuring a larger $N$ are less noisy than the lower resolution ones, as expected from Eq.~\ref{eq:r_Brown}; part of the oscillations in the trend of the CoM radius (especially at early times, and in the low-resolution cases) is due to the fact that the CoM orbital motion does not span a perfectly circular orbit, but exhibits some residual eccentricity. It is also important to notice that the final radius at which the MBHB CoM settles does not depend on the number of particles adopted in the run, supporting the fact that the CoM oscillations are not an effect of limited resolution (which instead plays a significant role in the traditional Brownian motion, Eq.~\ref{eq:r_Brown}).

\subsection{Effect of the MBHB center of mass motion on binary hardening}

In this section we explore the impact of the CoM oscillation on the MBHB hardening rate. This aspect is relevant as  the MBHB CoM wandering allows it to explore a region of space where it can interact with stars which otherwise would not be able to approach the binary. In this way, the binary  loss cone can be considered to be always full: the CoM oscillation may thus enhance the binary shrinking efficiency even for spherical systems in the collisionless limit.

\begin{figure}
  \centering
  \includegraphics[width=\hsize]{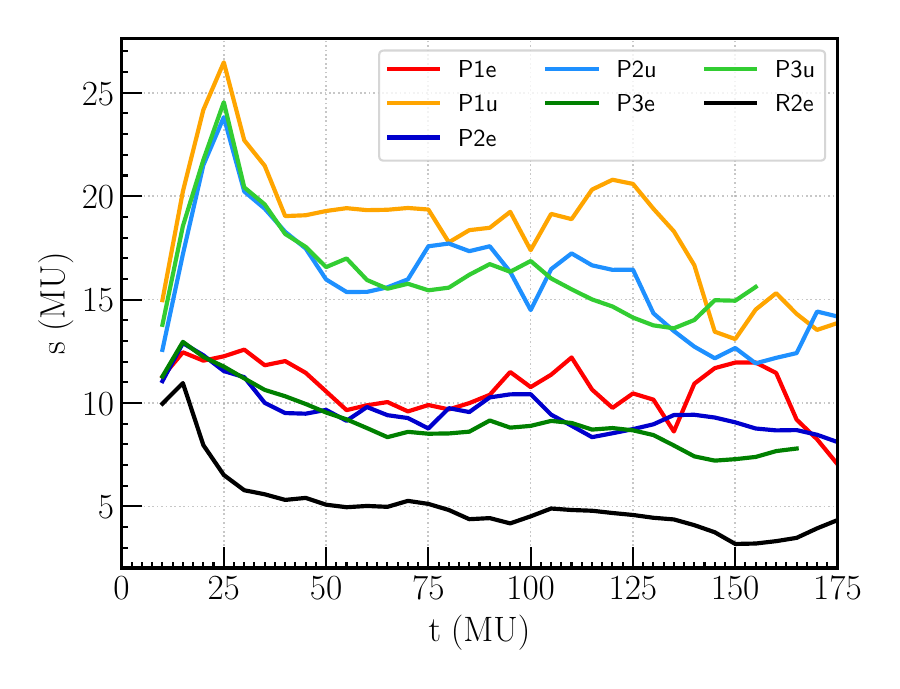}
    \caption{Time evolution of the hardening rates (Eq.~\ref{eq:s}) for the different runs presented in the paper. 
              }
    \label{fig:s}
\end{figure}

To quantify the efficiency at which the binary shrinks, it is customary to define the 
binary hardening rate $s$ as the time derivative of the inverse semi-major axis:
\begin{equation}\label{eq:s}
s= \frac{ \rm d}{{\rm d} t} \bigg (  \frac{1}{a}  \bigg).
\end{equation}
This quantity is a measure of the binary energy loss as a function of time.
Fig \ref{fig:s} shows the time evolution of the hardening rate for the presented runs, and  it is computed by fitting the slope of the inverse semi-major axis over short time intervals ($\Delta t = 1.25$).
The hardening rate evolution  for the prograde runs does not show a substantial dependence on the number of particles for each fixed mass ratio, and it stabilizes to $s\approx10$ ($s\approx 15$) for equal (unequal) mass binaries. On the other hand, the retrograde run (R2e) features a significantly smaller hardening rate (nearly a factor 2 smaller)  compared to the prograde equal mass runs. The fact that the retrograde run does not feature any oscillation about the centre apart from the traditional Brownian wandering, contrarily to the prograde cases, is an indication of the fact that the binary coherent oscillations ensure a more efficient loss cone refilling.

\begin{figure}
  \centering
  \includegraphics[width=\hsize]{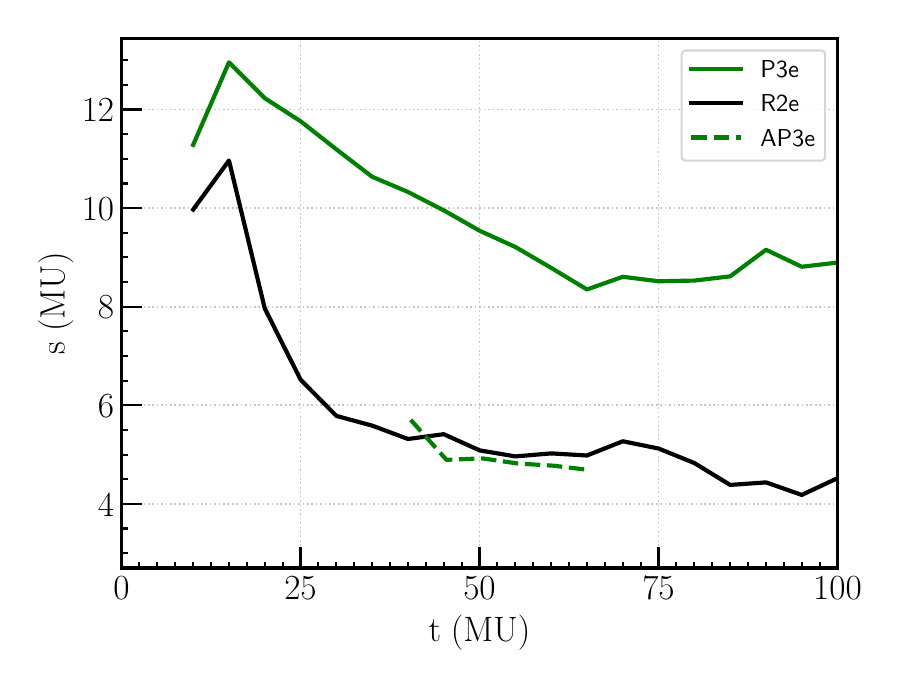}
    \caption{Time evolution of the hardening rates  (Eq.~\ref{eq:s})  for the prograde equal mass run P3e, the retrograde equal mass run R2e and the model AP3e, in which the CoM of the equal mass, prograde binary is fixed at the centre of the stellar distribution. If the binary is anchored in the centre, its hardening rate gets very similar to that of the retrograde run.
              }
    \label{s_counter}
\end{figure}

In order to have a deeper insight on the role of the binary oscillation on the loss-cone refilling, we performed a  run forcing the co-rotating binary in the P3e model to stay anchored to the system's center (A3Pe model), as   detailed in Sec.~\ref{sec:methods}. 
In Fig.~\ref{s_counter} the hardening rate of the anchored binary in AP3e is compared to that of the free co-rotating binary in the same resolution run, P3e, and of the counter rotating run, R2e. What emerges is that once the binary CoM orbital motion is inhibited, the binary hardening rate is nearly equal to that of the counter-rotating case. This is a very strong indication of the fact that the loss cone refilling within rotating systems hosting a prograde binary is induced by the MBHB CoM oscillation.

\subsection{CoM evolution for a single MBH}

\begin{figure}
  \centering
  \includegraphics[width=\hsize]{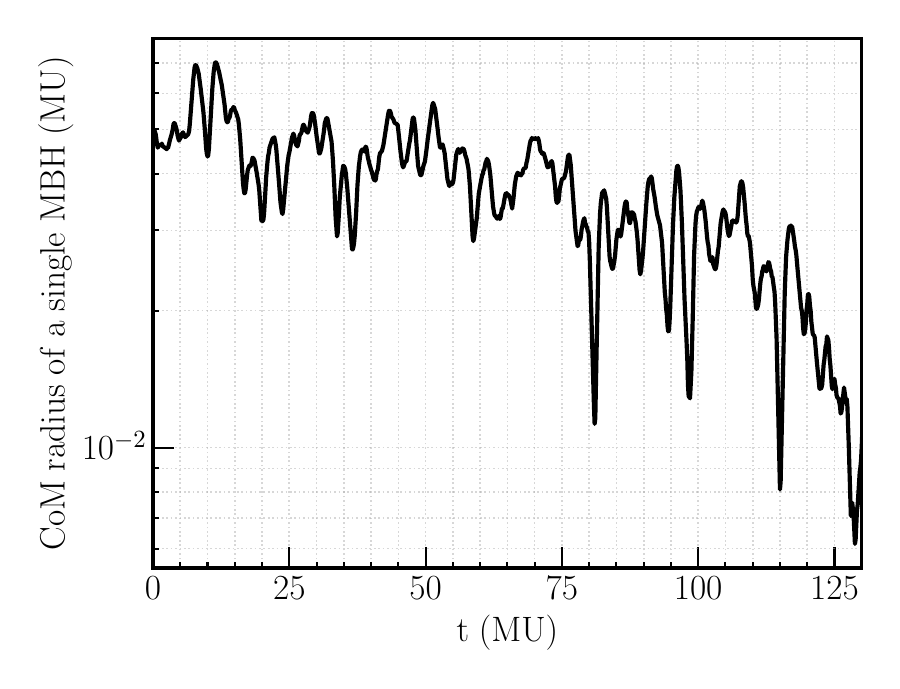}
    \caption{Displacement from the centre of the stellar distribution of a single MBH initialized by manually merging the binary in run P2e. Time $t = 0$ corresponds to the instant at which the MBHs in the progenitor binary are  merged. The MBH gradually inspirals towards the centre of the system in response to dynamical friction, and it does no longer exhibit coherent oscillations about the system centre. 
              }
    \label{R_t_merged}
\end{figure}

In order to better understand the nature of the MBHB wandering, and especially if slingshot interactions with passing stars are the responsible for the non-Brownian oscillation of prograde binaries, we perform an additional run in which we manually merge the MBHB in  model P2e  into a single MBH at time $t=70$. From this moment on, we track the displacement of the single MBH from the centre of the stellar distribution as a function of time. Fig.~\ref{R_t_merged} shows that, after the forced binary coalescence  the MBH gradually sinks back towards the center of the stellar distribution, and it stabilizes its oscillation radius to $\approx 0.01$ by $t\approx100$; the final radius nearly coincides with its expected Brownian wandering radius (see Eq.~\ref{eq:r_Brown} and Tab.~\ref{tab:Rbin}).\footnote{Note that the  Brownian wandering radius of a single MBH is expected to be nearly equal to the one of a binary with the same mass (Eq.~\ref{eq:r_Brown}, \citealt{Merritt01}).} 
This behavior is a strong indication of the fact that slingshot interactions with the binary sustain its CoM displacement and oscillation about the centre; once the binary has merged, the single MBH can sink back near the origin of the distribution as a result of dynamical friction.  This proves that single MBHs only experience the traditional Brownian wandering, regardless of the system rotation.
   

\section{Modelling of the CoM evolution}\label{sec:model}

\begin{figure}
  \centering
   \includegraphics[width=\hsize]{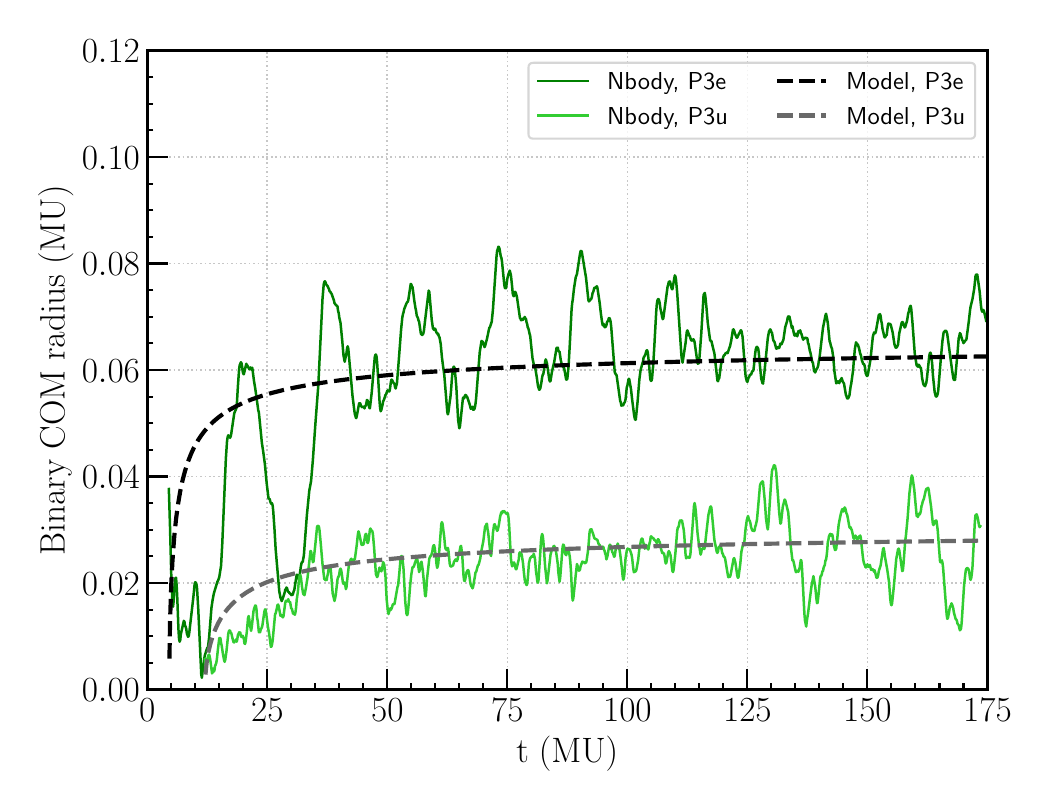}
      \caption{Time evolution of the binary CoM displacement from the centre of the system, $R_b$, as obtained from the simulations (solid lines) and from our theoretical modelling (Eq.~\ref{eq:model}). For model P3e we solved Eq.~\eqref{eq:model} assuming $\rho=0.8$, $\sigma =0.7$, $a_0=0.05$, $s=10$ and we initialize $R_b=0$ at $t=7.5$; for model P3u we set $\rho=2$, $\sigma=0.75$ $a_0=0.01$, $s=15$ and we initialize $R_b=0$ at $t=12$. These are characteristic values we extracted from the simulation. The most uncertaity is associated to the choice of $a_0$, as better detailed in the text and in footnote \ref{fn:model}.}
    \label{fig:model}
\end{figure}

To explain the behavior of the MBHB CoM in spherical rotating models, it is important to consider that, in the prograde scenario, virtually all stars approach the binary with a $z$ component of their angular momentum aligned with the binary angular momentum and typically larger than that of the binary, at least for the stages just after the binary formation, during which the binary external angular momentum experiences a significant growth. In addition, in the prograde runs, the binary eccentricity remains always very close to 0, or in other words, the MBHB has nearly the maximum angular momentum allowed for that given semi-major axis. At each prograde interaction, each star is thus likely to enhance the binary angular momentum. This enhancement can result in (i) an enlargement of the binary semi-major axis, but this almost never happens, as the interactions are typically found to shrink the binary (Fig.~\ref{bh_prop}); (ii) a reduction of the binary eccentricity, which is however already near its minimum, and it cannot decrease further; (iii) an enhancement of the external angular momentum of the binary, which is then the only viable option. 
In this situation, the time variation of the external binary angular momentum\footnote{Here we assume that the external binary angular momentum is aligned with the system rotation, as we find in our runs, and that the binary CoM orbital motion remains perfectly circular. } $L_{\rm ext}=M_b R_b v_b$, with $R_b$, $v_b$ radius and velocity of the binary CoM, should be equal to the rate at which the binary encounters stars times the typical angular momentum gained by the binary for each encounter. The stellar encounter rate can be written as  $dN/dt = 2\pi GM_{\rm b} a n_\star / \sigma$, where $a$ is the binary semimajor axis, while $n_\star$ and $\sigma$ respectively represent the stellar number density and velocity dispersion about the binary; the typical angular momentum exchange per stellar interaction is $\Delta L_\star\approx (m_\star / M_{\rm b})  L_{\rm int}$, where $L_{\rm int} = \mu \sqrt{GM_b a}$ is its internal angular momentum (in the -- verified -- assumption of a circular binary), and  $\mu$ is the reduced mass of the binary. It follows that 
\begin{equation}\label{eq:dldt}
    \frac{d L_{\rm ext}}{dt } = \frac{2\pi G\rho }{\sigma} \mu \sqrt{GM_b a^3},
\end{equation}
where $\rho=m_\star n_\star$. The CoM velocity $v_b$ is the circular velocity at the radius of the binary CoM; since the density profile remains nearly flat in the central region after the initial scouring, we can write
\begin{equation}\label{eq:vb}
    v_b=\sqrt{\frac {{4\pi G \rho}} 3}  R_b,
\end{equation}
i.e. the expected circular velocity at $R_b$; we checked the validity of this expression, and we found a very good match in our runs. On the right-hand side of Eq.~\ref{eq:dldt}, $a$ exhibits the strongest dependence on time (see e.g. $1/a$ in Fig.~\ref{bh_prop}): from  Eq.~\ref{eq:s} we can write
\begin{equation}\label{eq:at}
    a(t) = \frac{a_0}{1+a_0st},
\end{equation}
with $a_0=a(t=0)$.\footnote{Note that, in principle, this expression is valid only when the binary is hard, but for simplicity we assume it to be valid from the moment $R_b$ starts increasing; this is an approximation, but it is supported by the relatively limited variation of $s(t)$ in Fig.~\ref{fig:s}.} In this model we neglect the much milder time dependence of $\sigma$ (whose value within a radius of $\approx R_b$ only varies by nearly 10 per cent in our models) and $\rho$ (which nearly halves its value at $\approx R_b$ by the end of the integrations). Combining Equations~(\ref{eq:dldt}, \ref{eq:vb}, \ref{eq:at}) we obtain
\begin{equation}
    \frac{d}{dt} R_b^2 = \sqrt{\frac {3 \pi G^2 \rho } {\sigma^2} \frac{\mu^2}{M_b}  a^3(t)},
\end{equation}
whose solution reads, setting $R_b^2=0$ at $t=t_0$
\begin{align*}
R_b &= \sqrt{\frac{2 A}{B} \bigg(1- \frac{1}{\sqrt{1+B(t-t_0)}} \bigg)}  \numberthis \label{eq:model} \\ 
 A &= \sqrt{\frac {3 \pi G^2 \rho } {\sigma^2} \frac{\mu^2}{M_b}  a_0^3} \quad \quad
 B = a_0s;
\end{align*}
it is obviously valid only for $t\geq t_0$.

Fig.~\ref{fig:model} compares the evolution of $R_b$ in the simulations to what obtained from the above equation, for models P3e and P3u: our model seems to well reproduce the data. 
It is worth noting that the normalization of the curve in the plots (i.e., the value of $\sqrt{2A/B}$) is somewhat arbitrary, depending on the value one picks for the MBHB semimajor-axis $a_0$ at which $R_b$ starts growing.\footnote{\label{fn:model} Shortly after the binary formation (and in coincidence with the onset of the growth of $R_b$) the binary shrinks very quickly. Given the dependence of $R_b\propto a_0^{3/4}$, by picking different values of $a_0$ we obtain curves whose value gets larger or smaller by a factor of a few;  we believe this uncertainty is intrinsic in our simple treatment and we still believe our modelling can capture the evolution of $R_b$ to a decent degree.} This is due to the fact that the angular momentum exchange is proportional to the internal binary angular momentum, which is much larger near the binary formation time and strongly declines later. This also means that the interactions effectively displacing the binary from the centre are those occurring shortly after the binary formation time, while the ones occurring later impact less and less the external binary angular momentum evolution.

It is also worth accounting for the fact that dynamical friction should be acting on the binary CoM to bring it back to the centre, as it happens for the single MBH (Fig.~\ref{R_t_merged}). While in the beginning of the evolution the simulations clearly show that dynamical friction is subdominant compared to stellar interactions in inducing the evolution of $R_b$, this could be no longer true at later times. In order to check the relative importance of the two effects, we can compare the torque on the binary CoM on the right-hand side of Eq.~\ref{eq:dldt} to the torque we expect from dynamical friction.

However, the magnitude of dynamical friction in the present configuration cannot be trivially estimated, owing to the fact that the binary moves very close to the centre of a cored stellar distribution, in which fast moving stars may have an important contribution, and in which the estimate of the minimum and maximum impact parameter can be somewhat arbitrary. For this, we estimated the DF empirically, only focussing on the equal mass prograde runs. We start considering the time over which the single MBH of run P2u shown in Fig.~\ref{R_t_merged} is dragged back into the centre, given its initial angular momentum $L_{\rm ext} = M_b R_b v_b(R_b) \approx 6\times 10^{-5}$ (Tab.~\ref{tab:Rbin} and Eq.~\ref{eq:vb}), to write the associated dynamical friction torque as
$d L_{\rm DF}/{dt }\approx \Delta L_{\rm ext} / \Delta t \approx 5\times 10^{-7}$. This should be compared to the right hand side of Eq.~\ref{eq:dldt}, which can be rewritten, for the equal mass prograde cases, as
${d L_{\rm ext}}/{dt } \approx 1.8\times10^{-3}a^{3/2}$; this implies the two contributions to the evolution of the binary external angular momentum to be equal for $a\approx4.3\times10^{-3}$, and dynamical friction to be a factor 10 more efficient than stellar interactions at $a\approx9.2\times10^{-4}$. As a consequence, we expect that the binary should sink back towards the centre less than a hundred time units after the end of our prograde runs at $t\approx180$.\\

The model presented so far also allows to understand why the CoM does not undergo analogous oscillations in the retrograde scenario: in that case, stars can only deposit angular momentum that has opposite sign compared to the binary one, thus they  reduce the binary internal angular momentum instead of inducing a net oscillation in its CoM: this is supported by the fact that the eccentricity undergoes a continuous growth in the counter-rotating run (Fig.~\ref{bh_prop}). In principle, over sufficiently long timescales, the counter-rotating binary is expected to eventually flip the sign of its angular momentum and finally circularize \citep{Sesana11, Gualandris12}. However, since the external angular momentum growth occurs about the binary binding, and it is much less efficient at later times, we expect counter-rotating binaries to always remain close to the centre, even once they become prograde.


\section{Discussion and conclusion}\label{sec:discussion}

In this paper we tested the effect of spherical rotating stellar systems onto the dynamics of forming MBHBs.  
While we are perfectly aware that realistic rotating systems typically display some degree of flattening, we investigated rotating spherical systems as this allowed us to isolate the effect of rotation, avoiding additional effects possibly caused by the the global torques induced by deviations from spherical symmetry\footnote{Note that \cite{Holley-Bockelmann15} and \cite{Khan20} do indeed have flattened systems, but the rotation in their models is artificially introduced using our same procedure.}.

We found that prograde binaries (i.e. binaries with an angular momentum aligned with the net angular momentum of the stellar core) are forced out of the centre of their host galaxies due to the interaction with their background. The CoM of prograde binaries starts moving on quasi-circular orbits around the centre of the stellar core. Such motion is considerably larger than the typical Brownian wandering experienced by MBHBs evolving in isotropic backgrounds, and introduces a time-dependence in loss-cone of the binaries, that remains full during their whole shrinking. We demonstrated through dedicated numerical experiments that such results (the enhanced binary CoM wandering and the fast hardening rate) are not valid for retrograde binaries nor for single MBHs: indeed the artificial merger of a wandering prograde MBHB leads to the return of the MBH remnant to the centre of the system, demonstrating that the physical process driving the CoM motion is the energy and angular momentum exchange between (prograde) binaries and single stars.   

Our investigation improves upon the previous papers presenting the circling of the binary CoM and the binary enhanced hardening evolving in rotating axi-symmetric systems  \citep{Holley-Bockelmann15, Mirza17, Khan20} in two respects: (1) The deviations from spherical symmetry in the initial condition of such seminal investigations prevented a clear identification of the physical driver of the observed binary evolution. Indeed, in such geometries the global torques exerted by the whole stellar distribution onto single stars could play a role in the refilling of the loss-cones of the MBHBs \citep[but see][for a different point of view]{2015ApJ...810...49V}. With our simplified (spherical) stellar distribution we proved that rotation alone can cause both the MBHB circling and the boosted hardening observed; (2) we complemented our numerical study with a phenomenological analytical model that reproduces the evolution of the binary CoM observed in the prograde runs, strenghtening the proposed physical interpretation of the behavious observed in the simulations.

A remarkable difference between our results and those obtained by \citet{Holley-Bockelmann15} regards the hardening rates of retrograde binaries. In the rotating-spherical scenario we find that retrograde binaries shrink at a significantly slower pace than their prograde counterparts, while such difference is not observed in the rotating-flattened scenario discussed by Holley-Bockelmann and collaborators. In our analytical model the different behaviours are due to the absence of any binary CoM motion larger than the Brownian motion typically observed in isotropic systems, that prevent any significant collisionless loss-cone refilling associated to the motion of the binary CoM. The disagreement with the findings of \citet{Holley-Bockelmann15} could, in principle, be due to the different geometries of the stellar distributions, motivating further modeling of axi-symmetric systems.

Our analytical model and our numerical experiments agree on 
the fact that MBHBs experience the most external angular momentum growth right after their formation, at large semi-major axes. This implies that binaries forming with their internal angular momentum significantly offset from that of the surrounding environment would neither experience the CoM circling nor the enhanced hardening\footnote{But see the discussion above about the comparison with \cite{Holley-Bockelmann15}.}, as they would have shrunk their semi-major axis significantly before getting aligned with the environmental angular momentum. It is however possible that, in systems with a significant amount of rotation at large scales, the internal angular momentum of the forming binaries is already aligned with the angular momentum of the surrounding environment. Such configurations are expected even for initially strongly misaligned galaxy mergers, as (1) at large scale dynamical friction  onto rotating systems would act on the massive bodies dragging them towards a prograde, circular orbit \citep[e.g.][]{Dotti2006, Bonetti2020, Bonetti2021}, and (2) the same process can take place even at smaller scales immediately before the binary formation \citep{Mirza17, Khan20}. 

The relevance of the background rotation for the evolution of MBHBs depends ultimately on the typical  dynamical properties of their hosts. For light host galaxies hosting light MBHs $10^5-10^7$ \msun, in the mass range detectable by the forthcoming LISA mission, clear rotation is commonly observed at low redshift both at galactic  and sub-kpc scales \citep[e.g.][]{Kormendy2013}. It is yet unclear for which mass ratios and up to which redshift the same rotationally dominated structures are expected in galaxy mergers. Dedicated observational studies and detailed analyses of cosmologically motivated galaxy merger simulations are needed to properly gauge the impact of the presented results on the whole population of MBHBs.




\section*{Acknowledgements}

A.S. and E.B. acknowledge financial support provided under the European Union’s H2020 ERC
Consolidator Grant ``Binary Massive Black Hole Astrophysics'' (B Massive, Grant Agreement: 818691).
We acknowledge the CINECA Award N. HP10C4GJTF for the availability of high performance computing resources and support.

\section*{Data Availability Statement}
The data underlying this article will be shared on reasonable request to the corresponding author.




\bibliography{bibliography} 


\appendix

\bsp	
\label{lastpage}
\end{document}